\documentstyle[twocolumn,aps,epsfig]{revtex}
\begin{document}
\draft
\title{Real Time Radiative Corrections to Charged Particle Decay Laws}
\author{A. Widom and J. Swain}
\address{Physics Department, Northeastern University, Boston MA U.S.A.}
\author{Y.N. Srivastava }
\address{Physics Department \& INFN, University of Perugia, Perugia, Italy}

\maketitle

\begin{abstract}
The real time exponential decay laws for meta-stable charged particles 
are shown to require radiative corrections. The methods employed are 
well known to be valid for radiatively correcting Breit-Wigner line shapes. 
Radiative corrections contribute substantially to 
precision life time measurements of muons and pions when 
initially stopped in condensed matter.
\end{abstract}  


\narrowtext

\section{Introduction}

It is well known\cite{1} that the exponential decay law\cite{2}  
for the survival probability $P_0(t)$ of a meta-stable particle, 
\begin{equation}
P_0(t)\approx\exp(-\Gamma t),
\end{equation}
arises from an energy distribution\cite{3} which is nearly Lorentzian 
in shape 
\begin{equation}
dW_0(E)\approx \Big({\Gamma \over 2\pi}\Big)
\Big({dE\over (E-E_0)^2+(\Gamma /2)^2}\Big)\ ;
\end{equation}
i.e. 
\begin{equation}
P_0(t)=\Big|\int e^{-iEt/\hbar }dW_0(E)\Big|^2\ \ 
({\rm exponential\ decay}).
\end{equation}
It is also very well known that when charged particles are involved 
in a decay process, the Lorentzian energy distribution $dW_0(E)$ 
of the meta-stable particle must be radiatively corrected\cite{4}\cite{5}
\cite{6} to a new energy distribution $dW(E)$. The new distribution has a 
much more ``skewed'' line shape than that of a simple Lorentzian. It then 
follows that the survival probability also has a radiative correction 
in real time; i.e.  
\begin{equation}
P(t)=\Big|\int e^{-iEt/\hbar }dW(E)\Big|^2\ \ 
({\rm radiatively\ corrected}).
\end{equation}
Our purpose is to discuss the real time consequences of radiation in 
the decay of meta-stable charged particles.

Previous to this work, the notion that radiative corrections have 
important implications\cite{7} for observations of quantum noise in 
$\alpha $-decays and $\beta $-decays in heavy nuclei was theoretically 
developed\cite{8} for real time counting rates. 
Strong experimental evidence for quantum noise in nuclear 
$\beta $-decay counting rates has been reported\cite{9}\cite{10}. 

The time scales of quantum noise observations in nuclear physics are 
of the order of a few hours to a few days. These long times are still
much less than the nuclear life times, which in turn are of the order 
of a few years. For the application of radiative corrections to real time 
decay laws in high energy physics, e.g. for the weak decay  
\begin{equation}
\pi^+\to \mu^+ + \nu_\mu ,
\end{equation}
or the weak decay 
\begin{equation}
\mu^+\to e^+ +\nu_e +\bar{\nu }_\mu ,
\end{equation}
the available absolute measurement time scales are considerably reduced. 
Thus, the methods of detecting real time radiative corrections should 
be somewhat different from those methods used for nuclear weak decays.

In Sec.II, the radiatively corrected survival probability $P(t)$ will 
be calculated from the mean number of soft photons $dN(\omega )$ 
radiated into the bandwidth $d\omega $. The properties of the resulting 
distribution are discussed in Sec.III, in terms of the time dependent 
transition rate for decay $\gamma (t)$, defined as 
\begin{equation}
P(t)=\exp\Big(-\int_0^t\gamma (s)ds\Big).
\end{equation}
In Sec.III we illustrate the computation of the time dependent 
transition rate for a simple two body decay of a charged particle, 
e.g. Eq.(5). In Sec.IV we discuss the soft photon emission 
coupling strength as a function of photon frequency and the velocity of 
the produced charged particle both for insulators and conductors. 
We discuss the soft photon 
emission in both the  Bremsstrahlung and  Cerenkov regimes. In Sec. V, 
we find (for times {\em short} on the scale 
of inverse soft photon frequencies) that there exists a ``hot spot'' 
in the decay rate; i.e. the decay rate per unit time $\gamma (t)$ has 
a sharp peak for short times. In the long time limit, the decay rate 
settles down to $\Gamma $, which determines the intrinsic life time. 
In the vacuum, as well as in materials, the decay rates exhibit 
a long time inverse $t$ asymptotic time dependence which may be described by 
\begin{equation}
\gamma (t)= \Gamma +\Big({2\beta_0 \over t}\Big)+\ ...\ \ {\rm as\ }
t\to \infty.
\end{equation} 
In Eq.(8), the over all coupling strength $\beta_0 $ for the soft photons  
may be computed from the mean number of radiated photons $dN(\omega )$ 
in a bandwidth $d\omega $; It is 
\begin{equation}
\beta_0 =\lim_{\omega \to 0}\ \omega \Big({dN(\omega )\over d\omega}\Big).
\end{equation}
In the concluding Sec.VI, we discuss why the notion of 
radiative corrections to real time decay measurements in high energy 
physics is surely worthy of further experimental study. 

\section{Energy Distributions and Survival Probabilities}

Let $\Psi $ denote the internal wave function of an unstable charged 
particle in the center of mass frame. The energy distribution of the 
state is given by 
\begin{equation}
dW(E)=\big(\Psi , \delta (E-H)\Psi \big)dE.
\end{equation}
The survival amplitude for the state $\Psi $ is given by 
\begin{equation}
{\cal S}(t)=\big(\Psi , e^{-iHt/\hbar } \Psi \big),
\end{equation}
which is rigorously related to the energy distribution via 
\begin{equation}
{\cal S}(t)=\int e^{-iEt/\hbar } dW(E).
\end{equation}
The survival probability 
\begin{equation}
P(t)=|{\cal S}(t)|^2 
\end{equation}
is thus given by Eq.(4). If, in a two body decay of a charged particle, 
$d{\cal P}(\omega )$ is the probability of emitting soft photon 
radiation in the energy interval $\hbar d\omega $, then the radiatively 
corrected renormalization $dW_0(E)\to dW(E)$ is computed via the energy 
convolution 
\begin{equation}
\Big({dW(E)\over dE}\Big)=
\int \Big({dW_0(E-\hbar \omega)\over dE}\Big)d{\cal P}(\omega ).
\end{equation} 
If, during the decay, there are a mean number $\bar{n}_k$ of photons 
radiated into mode $k$ with Poisson statistics, then 
\begin{equation}
{d{\cal P}(\omega )\over d\omega }=
\sum_{\{n\}}
\Big\{
\prod_k \Big({\bar{n}_k^{n_k}e^{-\bar{n}_k}\over n_k!}\Big)
\Big\}\delta \big(\omega -\sum_k n_k \omega_k \big).
\end{equation}
Employing the generating function in the time domain 
\begin{equation}
\int_0^\infty e^{-i\omega t}d{\cal P}(\omega )=e^{-\chi(t)},
\end{equation}
implies the simplification 
\begin{equation}
\chi(t)=\int_0^\infty \big(1-e^{-i\omega t}\big)dN(\omega ),
\end{equation}
where $dN(\omega )$ is the mean number of photons radiated into a 
bandwidth $d\omega $; i.e. Eqs.(15) and (16) imply Eq.(17) with 
\begin{equation}
dN(\omega )=\Big(\sum_k \bar{n}_k \delta (\omega -\omega_k)\Big)d\omega .
\end{equation}

Furthermore, Eqs.(1),(3),(4),(14) and (16) imply 
\begin{equation}
P(t)=\exp\big(-\Gamma t-2\ \Re e\ \chi(t) \big).
\end{equation}
Eqs.(17) and (19) for the radiatively corrected survival probability 
are the central results of this section. Eqs.(7), (17) and (19) imply 
that the radiatively corrected transition rate per unit time $\gamma (t)$
as a function of time is related to the mean number of soft photons 
$dN(\omega )$ radiated into a bandwidth $d\omega $ via 
\begin{equation}
\gamma (t)=\Gamma +2\int_0^\infty \omega \sin (\omega t)dN(\omega ).
\end{equation}
The asymptotic Eq.(8) follows from Eqs.(9) and (20). 

\section{Computation of Transition Rates}

In order to compute $\big(dN(\omega )/d\omega )$, and thereby 
$\gamma (t)$, for the decay of a charged particle stopped in matter, 
we employ Schwinger's photon propagator method\cite{11}. For a  
charged particle moving on a path $C$, the self action due to virtual 
photons is given by 
\begin{equation}
S={e^2\over 2c}\int_C\int_C {\cal D}_{\mu \nu}(x-y)dx^\mu dy^\nu ,
\end{equation}
where ${\cal D}_{\mu \nu}(x-y)$ is the photon propagator in the 
condensed matter\cite{12} wherein the original charged particle was 
stopped. Employing the $k$-space representation 
\begin{equation}
{\cal D}_{\mu \nu}(x-y)=\int D_{\mu \nu }(k)e^{ik\cdot (x-y)}
\Big({d^4 k\over (2\pi )^4}\Big).
\end{equation}
with 
\begin{equation}
L^\mu (k)=\int_C e^{ik\cdot x}dx^\mu ,
\end{equation}
the action 
\begin{equation}
S={e^2\over 2c}\int L^\mu (k)D_{\mu \nu }(k) L^\nu (k)^*
\Big({d^4 k\over (2\pi )^4}\Big).
\end{equation}
One may compute the mean number of radiated photons using 
\begin{equation}
N=2\Im m\big(S/\hbar \big).
\end{equation}
With the fine structure constant
\begin{equation}
\alpha =\Big({e^2\over \hbar c}\Big),
\end{equation} 
we have 
\begin{equation}
N=\alpha \int \Im m\Big(L^\mu (k)D_{\mu \nu }(k) L^\nu (k)^*\Big)
\Big({d^4 k\over (2\pi )^4}\Big).
\end{equation}
If $v_i$ represents the initial four velocity of a particle (before a two 
body decay) and $v_f$ represents the final recoil four velocity of the 
produced charged particle after decay, then one easily obtains the usual 
expression\cite{4} for $L(k)$; i.e.
\begin{equation}
L(k)=i\Big\{
\Big({v_f\over k\cdot v_f}\Big)-\Big({v_i\over k\cdot v_i}\Big)
\Big\}.
\end{equation}

For the {\em vacuum} radiation distribution case  
\begin{equation}
D^{vac}_{\mu \nu }(k)=\Big({4\pi \eta_{\mu \nu}\over k^2-i0^+}\Big),
\end{equation}
so that Eqs.(37), (38) and (39) imply the well known\cite{4} {\em vacuum} 
radiated photon distribution 
\begin{equation}
d^3 N_{vac}({\bf k})=
\Big({\alpha d^3{\bf k}\over 4\pi^2c|{\bf k}|}\Big)
\Big\{
\Big({v_f\over k\cdot v_f}\Big)-\Big({v_i\over k\cdot v_i}\Big)
\Big\}^2.
\end{equation}
In the vacuum rest frame of the charged particle (before decay), we find 
the usual result, as in Eq.(18), 
\begin{equation}
dN(\omega )=\int \delta (\omega -c|{\bf k}|)d^3N_{vac}({\bf k})=
\beta \Big({d\omega \over \omega }\Big).
\end{equation}
For example, if in the rest frame of the charged particle 
(before decay), a single final state charge has velocity ${\bf v}$, 
then
\begin{equation}
\beta ({\bf v})=
\Big({\alpha \over \pi}\Big)
\Big\{
\Big({c\over |{\bf v}|}\Big)
\ln\Big({c+|{\bf v}|\over c-|{\bf v}|}\Big)-2
\Big\}.
\end{equation}
Eqs.(20), (31) and (32) yield
\begin{equation}
\gamma (t)=\Gamma +\Big({2\beta ({\bf v})\over t}\Big),
\ \ ({\rm vacuum\ decay}).
\end{equation}
The radiative corrections for a charged particle decay, when 
the particle has been stopped in condensed matter, are somewhat 
more subtle. 

\section{Decay of Particles Stopped in Condensed Matter}

Condensed matter systems are often described (for complex frequency 
$\zeta $ in the upper half $\Im m(\zeta )>0$  plane) by a dielectric response 
function $\varepsilon (\zeta )$, or by a conductivity response function 
$\sigma (\zeta )$. These are related by 
\begin{equation}
\epsilon(\zeta )=1+\Big({4\pi i\sigma (\zeta )\over \zeta  }\Big).
\end{equation} 
The above description may be incorporated into the temporal gauge 
photon propagator, $\zeta \to (|\omega |+i0^+)$, written\cite{12} as 
$$
{\bf D}({\bf k},\omega )=
$$
\begin{equation}
\Big({4 \pi \over |{\bf k}|^2-(\omega /c)^2
\varepsilon(|\omega |+i0^+)}\Big)
\Big\{
{\bf 1}-\Big({c^2 {\bf kk}\over \omega^2 
\varepsilon(|\omega |+i0^+)}\Big)
\Big\}.
\end{equation} 
Employing the temporal gauge Eq.(35) in the evaluation of the number 
of radiated photons in Eqs.(27) and (28) yields 
\begin{equation} 
\beta (\omega ,{\bf v})=
\omega \Big({dN(\omega ,{\bf v})\over d\omega }\Big),
\end{equation}
i.e. 
\begin{equation}
\beta (\omega ,{\bf v})
=\Big({\omega \alpha \over \pi c}\Big)\Im m \int 
\Big({{\bf v\cdot D}({\bf k},\omega +i0^+) {\bf \cdot v} 
\over ({\bf k\cdot v}-\omega )^2}\Big) 
\Big({d^3 {\bf k}\over (2\pi)^3 }\Big).
\end{equation}
The condensed matter version of the vacuum Eq.(32) is found (after 
some tedious integration) to be
$$  
\beta (\omega ,{\bf v})=
$$
\begin{equation}
\Re e
\Big\{
\Big({\alpha \over \pi \sqrt{\varepsilon(\omega +i0^+ )}}\Big)
{\cal F}\big(z(\omega +i0^+),z^*(\omega +i0^+)\big)
\Big\}
\end{equation}
where
\begin{equation}
z(\zeta )=\Big({c\over v\sqrt{\varepsilon(\zeta )}}\Big),
\end{equation}
\begin{equation}
{\cal F}(z,z^*)=\Big({{\cal G}(z)-{\cal G}(z^*)\over z-z^*}\Big)
\end{equation}
and 
\begin{equation}
{\cal G}(z)=\Big({z^2-1\over 2}\Big)\ln\Big({z+1\over z-1}\Big)-z.
\end{equation}
In the limit of real values for $z$
\begin{equation}
\lim_{y\to 0}{\cal F}(x+iy)=
\Big\{
x\ln\Big({x+1\over x-1}\Big)-2 
\Big\},
\ \ {\rm if\ }|x|>1.
\end{equation}

\begin{figure}[htbp]
\begin{center}
\mbox{\epsfig{file=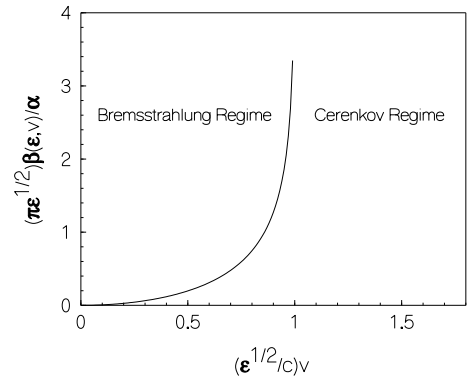,height=70mm}}
\caption{Bremsstrahlung coupling strength as a function of velocity for an 
insulator from Eqs.(38) and (42).}
\label{rfig1}
\end{center}
\end{figure}

In a (somewhat unphysical) model where the dielectric response 
$\varepsilon$ is independent of frequency, there will be a 
Bremsstrahlung regime for velocities obeying as shown in the above FIG.1;
\begin{equation}
|{\bf v}|<\Big({c\over \sqrt{\varepsilon }}\Big)
\ \ {\rm Brehmsstrahlung}.
\end{equation}
The value of the coupling strength in the insulating material 
Bremsstrahlung regime is given by 
\begin{equation}
\beta ({\bf v},\varepsilon )=
\Big({\alpha \over \pi \varepsilon }\Big)
\Big\{
\Big({c\over |{\bf v}|}\Big)
\ln\Big({c+ \sqrt{\varepsilon }|{\bf v}|
\over c- \sqrt{\varepsilon }|{\bf v}|}\Big)-2
\Big\}.
\end{equation}
In the high velocity regime there will be Cerenkov radiation. 
\begin{equation}
c>|{\bf v}|>\Big({c\over \sqrt{\varepsilon }}\Big)
\ \ {\rm Cerenkov}.
\end{equation}
However, the Cerenkov radiation regime can be discussed carefully only in 
models where the full complex dielectric response is taken into account.
The dissipation in physical continuous media implies a finite 
$\beta (\omega ,{\bf v})$ in all regimes. 

For example, in a model for an Ohm's law conducting material, 
the dielectric response function obeys 
\begin{equation}
\varepsilon (\omega )=1+\Big({4\pi i\sigma \over \omega }\Big)+...\ .
\end{equation}
For a non-relativistic particle the resulting coupling strength 
$\beta (\omega ,{\bf v})$ is plotted in FIG.2.

\begin{figure}[htbp]
\begin{center}
\mbox{\epsfig{file=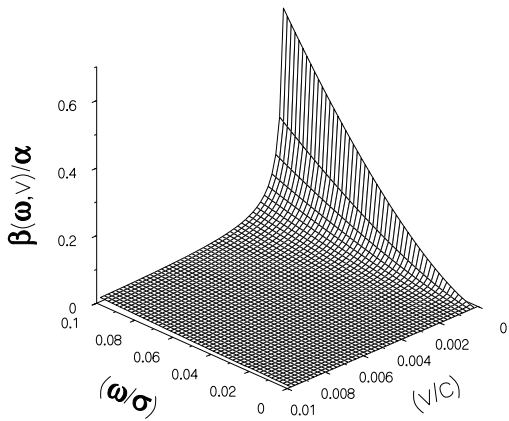,height=70mm}}
\caption{Coupling strength for a non-relativistic particle in a conductor.}
\label{rfig2}
\end{center}
\end{figure}

One may look at high velocity (for some values of 
$(\omega /\sigma)$) to note the almost discontinuous jump from 
the Bremsstrahlung regime to the Cerenkov regime in an Ohm's law 
metal. This is shown for $(\omega /\sigma)=0.4$ in FIG.3. For a 
high conductivity metal, $\sigma \sim 10^{18}/sec $. Thus, 
$\hbar \sigma \sim 1\ KeV$ which establishes the order of magnitude 
of the maximum frequency with which to define ``soft photons'' in 
this Ohm's law conducting model. 

Finally, for a conductor in the low frequency limit 
\begin{equation}
\Big({1\over \sqrt{\varepsilon }}\Big)\to \big(1-i\big)
\sqrt{\Big({\omega \over 8\pi \sigma }\Big)}, \ \ 
{\rm as}\ \ \omega\to 0,
\end{equation} 
and
\begin{equation}
z\to \big(1-i\big)\Big({c\over v}\Big)
\sqrt{\Big({\omega \over 8\pi \sigma }\Big)}, \ \ 
{\rm as}\ \ \omega\to 0.
\end{equation}.
From Eqs.(38), (40), (41), (47) and (48) it follows that 
\begin{equation}
\beta_0({\bf v})=\lim_{\omega \to 0}\beta (\omega ,{\bf v})
=\Big({\alpha |{\bf v}|\over 2c}\Big), \ \ ({\rm conductor}),
\end{equation}  
independent of the conductivity $\sigma $.

\begin{figure}[htbp]
\begin{center}
\mbox{\epsfig{file=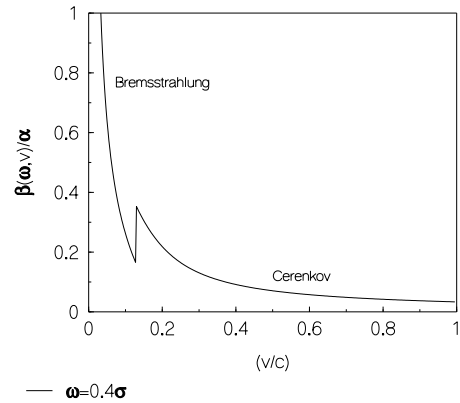,height=70mm}}
\caption{The almost discontinuous transition from the Bremsstrahlung to 
the Cerenkov regimes in a conductor.}
\label{rfig3}
\end{center}
\end{figure}

\section{Hot Spots and Long Time Tails}

A typical model for the mean number of photons $dN(\omega )$ in a 
bandwidth $d\omega $ employs an exponential cut-off for high frequency; 
it reads  
\begin{equation}
dN(\omega)=\beta_0 e^{-\omega \tau }\Big({d\omega \over \omega}\Big).
\end{equation}
where $(1/\tau )$ is the frequency cut-off. From Eqs.(20) and (50) it 
follows that 
\begin{equation}
\gamma (t)=\Gamma +2\beta_0\Big({t\over t^2+\tau ^2}\Big)
\end{equation}
may be used to compute the time dependent transition rate in Eq.(7). If 
$t>>\tau $, then the time dependence of $\gamma (t)$ is given by Eq.(8). 
Ultimately $\lim_{t\to \infty}\gamma (t)=\Gamma $, i.e. the intrinsic 
decay rate. For very short times the intrinsic rate also dominates; i.e.  
$\lim_{t\to 0}\gamma (t)=\Gamma $. However, for intermediate times there 
exists a peak or ``hot spot'' in which the decay rate may increase 
substantially. The situation is shown in FIG.4 which should hold true 
for both the vacuum decay and for decays in an insulator. For the insulating 
case $\beta_0$ is plotted in FIG.1. 

\begin{figure}[htbp]
\begin{center}
\mbox{\epsfig{file=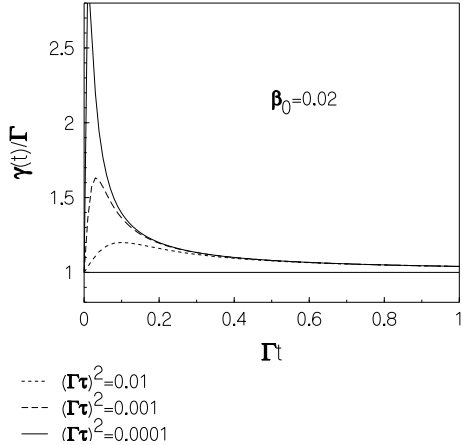,height=70mm}}
\caption{Transition rate ``hot spots'' for various cut-offs.}
\label{rfig4}
\end{center}
\end{figure}

By plotting $\gamma (t)$ in Eq.(7), one emphasizes the short time 
deviations from the uniform in time transition rate $\Gamma $. More 
conventionally\cite{13}\cite{14}, experimentalists directly plot 
$ln\ P(t)$. In such plots, the short time hot spot appears not merely less 
pronounced, but in reality {\em hardly visible}. Thus, it may have escaped 
some deserved attention.

\begin{figure}[htbp]
\begin{center}
\mbox{\epsfig{file=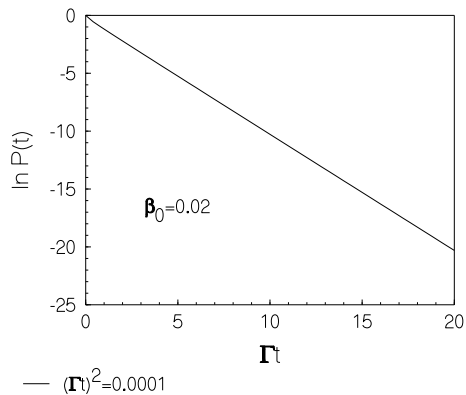,height=70mm}}
\caption{In a direct plot of ln P(t), the hot spot is not very prominent.}
\label{rfig5}
\end{center}
\end{figure}

From Eqs,(7) and (51), it is found in simple Bremsstrahlung models that 
\begin{equation}
P(t)=\Big({\tau^2\over t^2+\tau^2}\Big)^{\beta_0}exp\big(-\Gamma t)
\end{equation} 
To see what is involved, we plot the survival probability 
$P(t)$ in the above FIG.5. The short time hot spot for 
$(\Gamma \tau )^2=0.0001$, which is so very 
obvious when plotted as in FIG.4, is not at all so obvious when plotted as 
in Fig.5. Both curves are mathematically equivalent in accordance with 
Eq.(7). To explore the short time hot spot on an experimental level, 
one must examine the data in some detail during the period of the 
first life time of the survival probability, (say) at the early times 
$0<t<(0.5/\Gamma)$. For precision life time measurements, one tries a 
much wider time interval, (say) $0<t<(10/\Gamma)$. Such wide time intervals 
may mask important material effects which may have an effect on 
experimental precision. According to Eq.(52), the exponential decay law 
$exp(-\Gamma t)$ has a materials dependent prefactor
$\big(1+ (t/\tau )^2\big)^{-\beta_0}$ when explored over some 
twenty life times.

Eq.(52) holds true for both conductors and insulators when the produced 
charged particle is in the Bremsstrahlung regime. The Cerenkov regime 
for insulators is a bit more subtle. If one may define\cite{15}  
for an insulator, a Debye relaxation time $\tau_D$, via 
\begin{equation}
\Im m \big(\varepsilon (\omega +i0^+)\big)\to 
\varepsilon  (\omega \tau_D), \ \ 
{\rm as}\ \ \omega \to 0, 
\end{equation}
then in the Cerenkov regime $v>(c/\sqrt{\varepsilon})$, and in 
the low frequency limit $\omega \to 0$,
\begin{equation}
\beta_C (\omega ,|{\bf v}|)\to 
\Big({\alpha v\over c}\Big)\Big({1\over \omega \tau_D}\Big)
\Big\{1-\Big({c^2\over \varepsilon v^2}\Big)\Big\}.
\end{equation}
From Eqs.(20) and (54) it follows that in the insulating Cerenkov 
regime, the observed transition rate is renormalized to 
\begin{equation}
\gamma_C=\Gamma +\Big({\pi \alpha v\over c\tau_D}\Big)
\Big\{1-\Big({c^2\over \varepsilon v^2}\Big)\Big\}.
\end{equation}

Finally, the case of an insulator with a fractal low frequency 
behavior\cite{16} having exponent $\eta $,
\begin{equation}
\Im m \big(\varepsilon (\omega +i0^+)\big)\to 
\varepsilon \Big({\omega \tau_D \over |\omega \tau_D|^\eta }\Big),
\ \  {\rm as}\ \ \omega \to 0,
\end{equation}
leads to 
\begin{equation}
\gamma_C(t,\eta )=\Gamma + \varphi (\eta )\Big({\alpha v\over c\tau_D}\Big)
\Big\{1-\Big({c^2\over \varepsilon v^2}\Big)\Big\}
\Big({\tau_D\over t}\Big)^\eta ,
\end{equation}
where 
\begin{equation}
\varphi (\eta )=2\Gamma (\eta )\sin\Big({\pi \eta \over 2}\Big), \ \ 
\Gamma (\eta )=\int_0^\infty x^\eta e^{-x}\Big({dx\over x}\Big).
\end{equation}
From Eqs.(7) and (57) follows the Cerenkov fractal exponent survival 
probability 
\begin{equation}
P_C(t)=\exp\Big(-\Gamma t - \Phi_C (t,\eta ,\tau_D)\Big),
\end{equation}
where
\begin{equation}
\Phi_C (t,\eta ,\tau_D)=
\Big({\varphi (\eta ) \alpha v\over (1-\eta )c}\Big)
\Big\{1-\Big({c^2\over \varepsilon v^2}\Big)\Big\}
\Big({t\over \tau_D}\Big)^{(1-\eta )},
\end{equation}
and which exhibits (for short times) a stretched exponential form.
Note as $\eta \to 0$, Eq.(60) becomes equivalent to Eq.(55).

\section{Concluding Remarks}

Deviations from the exponential laws in real time triggered by
soft photons have been studied for decaying charged particles 
in materials. Under the {\em standard assumption} of 
factorizability of the ``dynamical'' energy distribution from the 
soft photon emission spectrum as given in Eq.(14), we have derived 
the transition rates for decays in insulating and conducting
materials, both in the Bremsstrahlung and Cerenkov regimes. Several 
interesting results emerge some of which are listed below. 

If a particle decays in a conducting material, a novel 
discontinuity is predicted to occur. We show in Fig. 3, that as the 
velocity of the charged particle produced in the decay increases from 
low values (bremsstrahlung region), the radiative coupling strength 
(and hence the transition rate) first decreases rapidly, shows a sharp 
discontinuity as it enters the Cerenkov region and then continues to 
decrease more slowly. Much care would be needed to experimentally 
observe such a discontinuity since it is rather sharp and hence 
confined to a very narrow velocity range of the produced particle.

Another prediction concerns ``hot spots'' and long 
time tails occurring both for the vacuum as well as an insulator. As 
exhibited in Fig.4, the transition rate has a well defined maximum
(hot spot) for intermediate times and is substantially different from
its asymptotic value. On the other hand, the same effect is shown
in Fig.5 to be completely washed out in a standard logarithmic plot 
of the survival probability data commonly presented by experimentalists. 
Thus, evidence, if any, for such an effect must be sought out through 
a careful study of the transition rate in the short time interval, say 
$0<t<(0.5/\Gamma)$. 

The Cerenkov regime for insulators is also of particular interest since
it leads to decays of the ``stretched exponential'' form. The
fractal exponent in the absorptive part of the low frequency limit of
the dielectric constant is shown to be directly related
to the radiative exponent in the real time decay [see Eqs.(56-60)]. It 
would be worthwhile to check it experimentally.

In view of the above predicted deviations from purely 
exponential decays, we urge that a concentrated, systematic and precise 
experimental study of the transition rate  be undertaken, 
both in the vacuum as well as in diverse materials for
different decay particle speeds. Exponential decays and Poisson 
statistics are almost axiomatic in experimental particle physics. Thus, 
any deviations are surely of fundamental interest.


\begin{thebibliography}{99}

\bibitem{1} V. F. Weisskopf and E. P. Wigner, {Z. Phys.} {\bf 63}, 
54 (1930).
\bibitem{2} M. L. Goldberger and K.M. Watson, {\it Collision Theory}, 
Wiley, New York (1964).
\bibitem{3} R. G. Newton, {\it Scattering Theory of Waves and Particles}, 
McGraw Hill, New York (1966)
\bibitem{4} E. Etim, G. Pancheri and B. Touschek, {\it Il Nuovo Cimento} 
{\bf 51B}, 276 (1967).
\bibitem{5} G. Pancheri, {\it Il Nuovo Cimento} {\bf 60}, 321 (1969). 
\bibitem{6} M. Greco, G. Pancheri and Y. N. Srivastava, 
{\it Nuc. Phys.} {\bf B101}, 234 (1975). 
\bibitem{7} P. H. Handel, {\it Phys. Rev.} {\bf A22}, 745 (1980).
\bibitem{8} C. M. Van Vliet and P. H. Handel, {\it Physica} 
{\bf 113A}, 261 (1982).
\bibitem{9} M. A. Azhar and K. Gopala, {\it Phys. Rev.} {\bf A39}, 
4137 (1989).
\bibitem{10} M. A. Azhar and K. Gopala, {\it Phys. Rev.} {\bf A44}, 
1044 (1991).
\bibitem{11} J. Schwinger, {\it Magnetic Charge} in {\it Proc. Conf. on 
Gauge Theories and Modern Field theory}, Northeastern University, 
Ed. R. Arnowitt and P. Nath, M.I.T. Press, Cambridge MA (1976).
\bibitem{12} A. A. Abrikosov, L.P. Gorkov and I. E. Dzyaloshinski, 
{\it Methods of Quantum Field Theory in Statistical Physics}, Chapt. 6, 
p 258, Dover, New York (1975). 
\bibitem{13} M. P. Balandin, V. M. Grebenyuk, V. G. Zinov, A.B. Conin 
and A. M. Ponomarev {\it JETP} {\bf 40}, 811 (1974).
\bibitem{14} G. Bardin, J. Duclos, A. Magnon, J. Martino, E. Zavatini, 
A. Bertin, M. Capponi, M. Piccinini and A. Vitale, {\it Phys. Lett} 
{\bf B 137}, 135 (1984).
\bibitem{15} H. Frohlich, {\it Theory of Dielectrics}, p. 70 
Clarendon Press, Oxford (1981). 
\bibitem{16} B. K. P. Svarife, {\it Principles of Dielectrics}, p. 66 
Clarendon Press, Oxford (1987).     

\end{thebibliography}
\end{document}